\documentclass[conference]{IEEEtran}
\IEEEoverridecommandlockouts
\usepackage{amsmath,amssymb,amsfonts}
\usepackage{algorithmic}
\usepackage{graphicx}
\graphicspath{{./figs/}}
\usepackage{bm}
\usepackage{cite}
\usepackage{comment}
\usepackage{xcolor}
\usepackage{caption,subcaption}
\usepackage{float}
\usepackage{xcolor}
\usepackage[utf8]{inputenc}
\usepackage{booktabs} 

\def\BibTeX{{\rm B\kern-.05em{\sc i\kern-.025em b}\kern-.08em
		T\kern-.1667em\lower.7ex\hbox{E}\kern-.125emX}}
\begin{document}
	
	\title{Performance Analysis of Communication Signals for  Localization in Underwater Sensor Networks
		\thanks{This work was partially supported by the Wallenberg AI, Autonomous Systems and
			Software Program (WASP) funded by the Knut and Alice Wallenberg Foundation.}
	}
	\author{\IEEEauthorblockN{Ashwani Koul}
		\IEEEauthorblockA{\textit{Division of Automatic Control,} \\
			\textit{Linköping University,}\\
			Linköping, Sweden. \\
			ashwani.koul@liu.se}
		\and
		\IEEEauthorblockN{Gustaf Hendeby}
		\IEEEauthorblockA{\textit{Division of Automatic Control,} \\
			\textit{Linköping University,}\\
			Linköping, Sweden. \\
			gustaf.hendeby@liu.se}
		\and
		\IEEEauthorblockN{ Isaac Skog}
		\IEEEauthorblockA{\textit{Division of Communication Systems,} \\
			\textit{KTH Royal Institute of Technology,}\\
			Stockholm, Sweden. \\
			skog@kth.se}
	}
	
	\maketitle
	
	\begin{abstract}
		Efficient localization in underwater sensor networks faces challenges due to limited bandwidth, energy constraints, and hardware complexity. Traditional systems separate sensing and communication, often resulting in inefficient resource usage. To address this, integrated sensing and communication (ISAC) has emerged, leveraging shared waveforms for both functions. This paper investigates the feasibility of using communication-centric waveforms for underwater localization. Specifically, we evaluate the performance of super-permutated frequency shift Keying and multiple frequency shift keying signals using a Cramér–Rao lower bound framework in a simplified bistatic scenario. Simulations incorporate temporally correlated auto-regressive AR(1) noise and varying signal-to-noise ratio levels to assess localization accuracy. A comparative analysis with a traditional sonar waveform, linear frequency modulated signal, highlights the potential of communication signals for dual-purpose ISAC applications in bistatic sonar configurations in underwater environments.
	\end{abstract}

	\begin{IEEEkeywords}
		Communication signals, Cramér–Rao lower bound, Fusion, ISAC, underwater target localization.
	\end{IEEEkeywords}
	
	\section{Introduction}
	
	Accurate localization of underwater targets is essential for applications such as maritime surveillance, autonomous navigation, and environmental monitoring~\cite{Bosser2025}. Underwater wireless sensor networks enable distributed sensing and communication among sensor nodes, allowing for cooperative localization using passive and active measurements~\cite{Akyildiz2005}. 
	
	Traditionally, underwater sensing and communication are treated as distinct functionalities, requiring separate waveform designs, signal processing chains, and hardware components. This separation introduces significant inefficiencies, particularly in underwater environments where resources are inherently constrained~\cite{Abraham2019}. Hence, the need to transmit separate signals for sensing and communication leads to suboptimal use of spectral and energy resources.
	
	To address this, integrated sensing and communication (ISAC) aims to unify sensing and communication functions by leveraging shared waveforms, signal processing, and hardware resources~\cite{weiintegrated23,liujrc20}. By combining these functions, ISAC addresses key limitations of traditional dual-system designs. For instance, in RF systems, ISAC waveforms can concurrently transmit data and process reflected signals for target detection, eliminating the need for separate radar and communication subsystems~\cite{Keskin2024}. 
	
	For underwater applications, the potential of ISAC remains largely underexplored. Prior studies have investigated communication-centric~\cite{jun2018detection, Men2022} and hybrid waveforms~\cite{niu2022integrated,yin2020integrated, Wang2023, Wang2024} for simultaneous data transmission and target detection, demonstrating the feasibility of reusing communication signals for sensing tasks. In~\cite{Jehangir2024,Liu2016}, the localization performance has been evaluated in monostatic setups using non-communication signals within ISAC frameworks. However, the potential of using communication waveforms in bistatic configurations, where spatially separated transmitters and receivers enable the reception of both direct-path and reflected signals, remains underexplored. Such configurations offer enhanced spatial resolution and richer geometric diversity by combining active and passive measurements, which can significantly improve localization accuracy.
	
	To analyse the localization efficacy, we evaluate the Cramér–Rao lower bound (CRLB) for joint passive and bistatic measurements in a simplified underwater scenario. We focus on two robust communication waveforms: super permutated frequency shift keying (SPFSK) and multiple frequency shift keying (MFSK), both designed to be resilient to multipath and Doppler effects. Additionally, we evaluate the wideband ambiguity function (WBAF) of each waveform to assess its resolution in delay and Doppler. A comparison with a traditional sonar waveform, the linear frequency modulated (LFM) signal, further demonstrates the localization potential of communication-centric signals in ISAC systems.
	
	To summarize, the key contributions of this paper are:
	\begin{itemize}
		\item A CRLB framework quantifying localization accuracy and guiding waveform design in an ISAC-enabled underwater sensor network.
		\item A performance evaluation of SPFSK and MFSK waveforms using CRLB and WBAF analysis, demonstrating their suitability for ISAC in underwater sensor networks.
	\end{itemize}
	
	\section{Methodology}\label{Prob_stat}
\begin{figure}[t]
		\centering
		\includegraphics[width=0.45\textwidth]{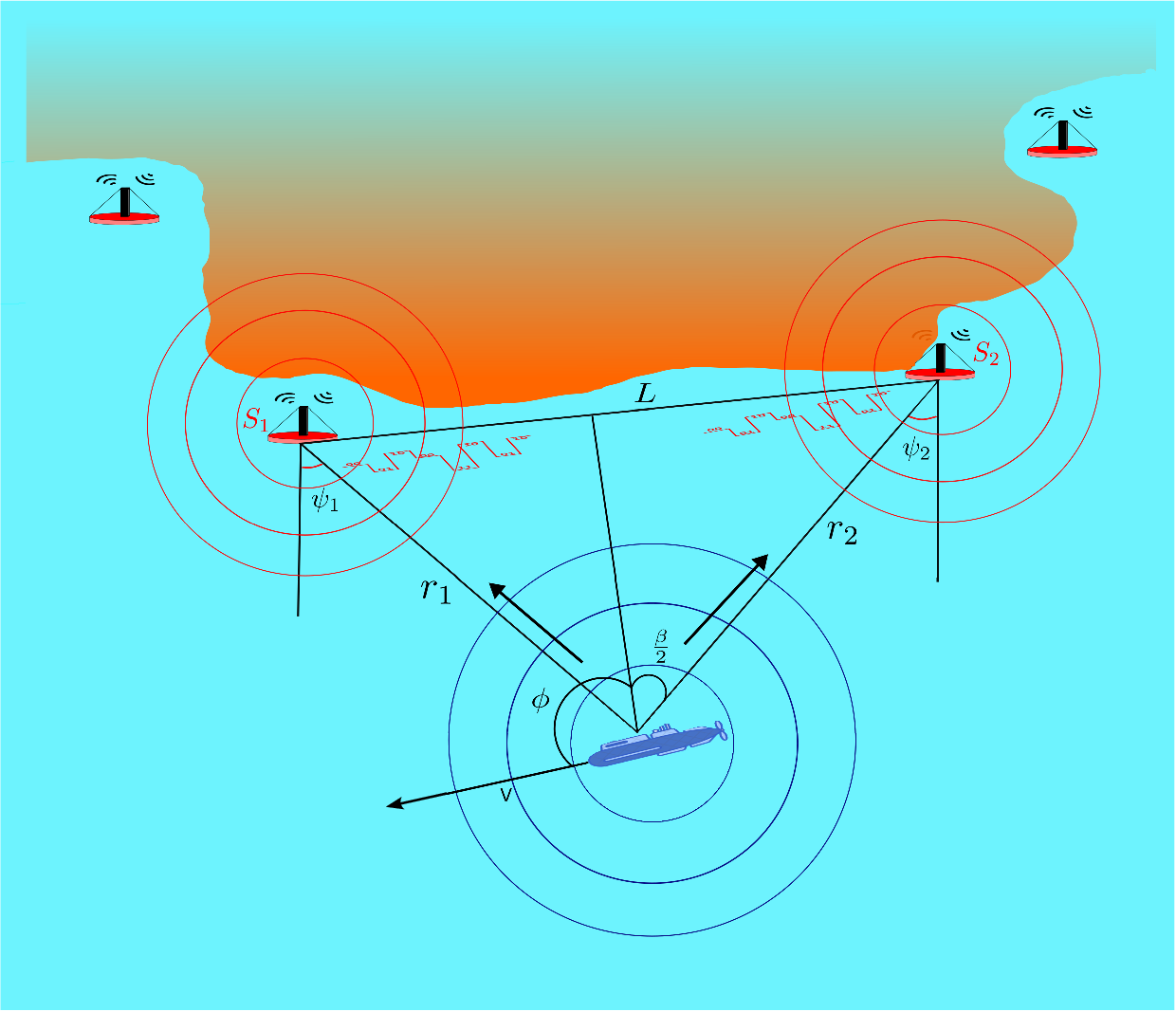}
		\caption{Underwater wireless sensor node configuration.}
		\label{Fig1}
	\end{figure}

	To investigate the feasibility of communication signals for localization, we consider a simplified underwater environment as shown in Fig.\ref{Fig1}. We investigate a scenario in which a sensor node passively listens for a target. Upon detection, the bearing estimates are communicated to other sensor nodes. The communication signal is received through both a direct path and a target-reflected path, along with the passive acoustic signal and ambient noise. The CRLB will be used to evaluate the theoretical achievable localization accuracy for the considered system setup.
	\subsection{Assumptions}
	To model the received signal and develop a CRLB framework, the following assumptions are made.
	\begin{itemize}
		\item The sensor nodes are each equipped with an $M$ element uniform linear receiver array.
		\item The locations of the sensor nodes are known, and they are time synchronized.
		\item The noise is assumed to be spatially white and temporally correlated. 
		\item The communication signal is correctly decoded at the receiver via the direct path.
		\item The target is an autonomous underwater vehicle (AUV) weighing $w$ [tons] and moves at a speed of $v$ [m/s]. 
	\end{itemize}
	\subsection{Signal Model}
	The $k^{\text{th}}$ batch of signal samples observed by the hydrophone array at the
	$i^{\text{th}}$ sensor node can be modeled as
	\begin{equation}\label{sensor_mm}
		\mathbf{y}_k^{i} =  \mathbf{u}_k^{(j,i)}(\bm{\theta}) +\mathbf{s}_k^{i}(\bm{\theta}) + \mathbf{e}_k^{i} \quad \in \mathbb{R}^{MN_{k}\times 1}
	\end{equation}
	where $N_k$ denotes the number of time samples observed during the $k^{\text{th}}$ time slot, $ \bm{\theta} = \begin{bmatrix}\mathbf{p}^\top & \eta\end{bmatrix}^\top$ denotes the unknown signal parameters to be inferred, with $\mathbf{p}=\begin{bmatrix} x & y\end{bmatrix}^\top$ and $\eta$ representing the target position and Doppler scaling factor respectively.
	Here, $\mathbf{u}_k^{(j,i)}(\bm{\theta})$ denotes the signal transmitted by the $j^{\text{th}}$ node and received at the $i^{\text{th}}$ node. If no signal is transmitted within the $k^{\text{th}}$ data batch, then $\mathbf{u}_k^{(j,i)}(\bm{\theta})=\mathbf{0}$.
	Moreover, $\mathbf{s}^{i}_k(\bm{\theta})$ and $\mathbf{e}_k^{i}$ denote the signal radiated by the target and the ambient noise, respectively. These signals are assumed to be random and independent, and distributed as
	\begin{align}
		\mathbf{s}_k^{i}(\bm{\theta}) &\sim \mathcal{N}\left(\mathbf{0},\,\mathbf{R}_{s,k}^{i}(\bm{\theta})\right) \\
		\mathbf{e}_k^{i}  &\sim \mathcal{N}\left(\mathbf{0},\,\mathbf{R}^i_{e,k}\right)
	\end{align}
	Here, $\mathbf{R}_{s,k}^{i}(\bm{\theta})$ and $\mathbf{R}^{i}_{e,k}$ denote the covariance structure of the signal radiated by the target and the ambient noise during the $k^{\text{th}}$ batch, respectively. Information about the structure of $\mathbf{R}_{s,k}^{i}(\bm{\theta})$ can be found in \cite{Bosser2025}. Additional information about the structure of measurement model in~\eqref{sensor_mm} and $\mathbf{R}^{i}_{e,k}$ is presented in Appendix~\ref{appendix1}.
	
	With reference to Fig. \ref{Fig1}, assume that during time slot $k-1$, no signal is transmitted in the network. However, both sensor nodes $S_1$ and $S_2$ observe the sound radiated by the target. In the next time slot, i.e., at time slot $k$, sensor node $S_1$ communicates the observed information to sensor node $S_2$. The joint signal model for all these batches of signal samples can then be defined as
	\begin{equation}\label{compact_pact}
		\mathbf{z} = \begin{bmatrix}
			\mathbf{y}_k^{S_2} \\
			\mathbf{y}_{k-1}^{S_1} \\
			\mathbf{y}_{k-1}^{S_2}
		\end{bmatrix} \sim \mathcal{N} \left(\bm{\mu}(\bm{\theta}),\bm{\Sigma}(\bm{\theta})\right)
	\end{equation}
	where
	\begin{equation}\label{mu}
		\bm{\mu}(\bm{\theta})  =
		\begin{bmatrix}  
			\mathbf{u}^{S_1,S_2}_k(\bm{\theta}) \\
			\mathbf{0} \\
			\mathbf{0}
		\end{bmatrix}
	\end{equation}
	and
	\begin{align}\label{sigma}
		\bm{\Sigma}(\bm{\theta}) =\begin{bmatrix}
			\bm{\Sigma}_{S_2,k}(\bm{\theta}) & \mathbf{0}  & \mathbf{0} \\
			\mathbf{0} & \bm{\Sigma}_{S_1,k-1}(\bm{\theta})  & \mathbf{0} \\
			\mathbf{0} &  \mathbf{0} & \bm{\Sigma}_{S_2,k-1}(\bm{\theta}) 
		\end{bmatrix}
	\end{align}
	with
	\begin{subequations}
		\begin{align}
			\bm{\Sigma}_{S_2,k}(\bm{\theta}) & = \mathbf{R}^{S_2}_{s,k}(\bm{\theta}) + \mathbf{R}^{S_2}_{e,k}, \\
			\bm{\Sigma}_{S_1,k-1}(\bm{\theta}) & = \mathbf{R}^{S_1}_{s,k-1}(\bm{\theta}) + \mathbf{R}^{S_1}_{e,k-1},\\
			\bm{\Sigma}_{S_2,k-1}(\bm{\theta}) & = \mathbf{R}^{S_2}_{s,k-1}(\bm{\theta}) + \mathbf{R}^{S_2}_{e,k-1}. 
		\end{align}
	\end{subequations}
	describing the covariance for the received signal at sensor nodes $S_1$ and $S_2$ during the respective batch.

	\subsection{Performance Metrics}
	Given the joint signal model in \eqref{compact_pact}, we assess the localization performance by computing the theoretical lower bound on the estimation error. This lower bound is obtained by first evaluating the Fisher information matrix (FIM) \cite{KayI}, which quantifies the amount of information that the observed data carries about the unknown parameters. The FIM for the $(q,r)$ entry is defined as
	\begin{align}\label{FIM_basic}
		\left[\mathcal{I}_{\bm{\theta}}\right]_{q,r} =& \left[ \frac{\partial \bm{\mu}^\top(\bm{\theta})}{\partial \theta_{q}} \bm{\Sigma}^{-1}(\bm{\theta}) \frac{\partial \bm{\mu}(\bm{\theta})}{\partial \theta_{r}}\right] \notag \\
		& + \frac{1}{2}\mathrm{tr}\left\{\bm{\Sigma}^{-1}(\bm{\theta}) \frac{\partial \bm{\Sigma}(\bm{\theta})}{\partial \theta_{q}} \bm{\Sigma}^{-1}(\bm{\theta}) \frac{\partial \bm{\Sigma}(\bm{\theta})}{\partial \theta_{r}}\right\},
	\end{align}
	where $\bm{\mu}(\bm{\theta})$ and $\bm{\Sigma}(\bm{\theta})$ are defined in \eqref{mu} and \eqref{sigma} respectively. 
	According to the Cramér–Rao inequality \cite{KayI}, the inverse FIM $\mathcal{I}_{\bm{\theta}}^{-1}$ lower bounds the covariance of an unbiased estimator. 
	Thus, to analyze the localization accuracy for the available measurements, we evaluate the root mean squared error lower bound (RCRLB) for the position $\mathbf{p}$ and Doppler scaling factor $\eta$, which are defined as
	\begin{subequations}
		\begin{align}
			\mathrm{R}\mathrm{CRLB}_{\mathbf{p}} &\triangleq \sqrt{ \sum_{q=1}^{2} \left[ \mathcal{I}_{\bm{\theta}}^{-1}\right]_{q,q}  } \label{eq:PEB} \\
			\mathrm{R}\mathrm{CRLB}_\eta &\triangleq \sqrt{ \left[ \mathcal{I}_{\bm{\theta}}^{-1} \right]_{3,3} } \label{eq:DEB}
		\end{align}
	\end{subequations}

	These lower bounds are evaluated numerically by modeling the mean $\bm{\mu}(\bm{\theta})$, and the covariance matrix $\bm{\Sigma}(\bm{\theta})$ under the condition that the radiated signal power and the ambient noise power are known.
	
	The overall FIM in~\eqref{FIM_basic} is computed using the combined measurement model defined in \eqref{compact_pact}, where the total covariance matrix $\bm{\Sigma}(\bm{\theta})$ has a block-diagonal structure. This structure arises from the independence of measurements across sensor nodes and time slots. As a result, the FIM is additive across observation blocks and can be expressed as
	\begin{equation}\label{FIM_indiv}
		\mathcal{I}_{\bm{\theta}} = \tilde{\mathcal{I}}^{(S_2,k)}_{\bm{\theta}} + \tilde{\mathcal{I}}^{(S_1,k-1)}_{\bm{\theta}} +\tilde{\mathcal{I}}^{(S_2,k-1)}_{\bm{\theta}}
	\end{equation}
	where $\tilde{\mathcal{I}}_{\bm{\theta}}^{(S_2,k)},\tilde{\mathcal{I}}_{\bm{\theta}}^{(S_2,k-1)}$ are the FIMs defined for the signal received at the $S_2$ sensor node during the $k^{\text{th}}$ and $k-1$ batch respectively, and $\tilde{\mathcal{I}}_{\bm{\theta}}^{(S_1,k-1)}$ is the FIM for the signal received at the $S_1$ sensor node for the $k-1$ batch.

	\subsection{SNR Equations}
	In any underwater environment, the signal energy decays due to the spreading, multipath propagation, and frequency-dependent attenuation of the signal. The dissipation also depends on the type and composition of the underwater channel. In the context of the signal models in (\ref{sensor_mm}), it is important to account for the degradation of signal energy caused by the underwater channel. Thus, the signal-to-noise ratio (SNR) for both the active and the passive received signal is modeled as
	\begin{align}\label{SNR_active}
		\mathrm{SNR}_{\text{active}}(r) & =\mathrm{SL}_{\text{active}} + \mathrm{TS} - \mathrm{PL}_1(r) -\mathrm{PL}_2(r) \notag \\
		&-(\mathrm{SL}_{\text{passive}}-\mathrm{PL}_2(r))- \mathrm{NL} \quad [\mathrm{dB}]
	\end{align}
	and
	\begin{equation}\label{SNR_passive}
		\mathrm{SNR}_{\text{passive}}(r) = \mathrm{SL}_{\text{passive}}-\mathrm{PL}_{i}(r)-\mathrm{NL}\quad [\mathrm{dB}] 
	\end{equation}
	where $\mathrm{SL}_{\text{active}}$ and $\mathrm{SL}_{\text{passive}}$ are the power levels (in dB re $\mu$Pa$^2$m$^2$) of the transmitted communication signal and the signal radiated by the target, respectively. $\mathrm{PL}_{i}(r), \forall i=\{1,2\}$ are the propagation losses (in dB re m$^2$) for the signal traveling between the sensor node and the target. $\mathrm{TS}$ represents the strength of the target (in dB re m$^2$) to reflect the incoming signal, and $\mathrm{NL}$ is the ambient noise power level (in dB re $\mu$Pa$^2$)\cite{Urick2013, Abraham2019}. For the case of $\mathrm{SNR}_{\text{active}}$, the signal component received in addition to the reflected communication signal is considered an interfering component and thus contributes to the overall noise floor.

	\subsection{Considered Scenarios}
	In underwater environments, the reflected signal from the target may not always be detected at the receiver node due to severe multipath, attenuation, or target scattering loss and high ambient noise levels. In such cases, the active bistatic signal is not detected, and localization relies solely on passive sensing. 
	
	To assess the impact of the bistatic signal's availability, we evaluate the RCRLB under two scenarios:
	
	\begin{itemize}
		\item  When the reflected signal is detected, bearing information from spatially distributed sensor nodes is combined with the information from the bistatic signal. 
		\begin{equation}\label{FIM_1}
			\mathcal{I}_{\bm{\theta}}  = \tilde{\mathcal{I}}^{(S_2,k)}_{\bm{\theta}} + \tilde{\mathcal{I}}^{(S_1,k-1)}_{\bm{\theta}} +\tilde{\mathcal{I}}^{(S_2,k-1)}_{\bm{\theta}}
		\end{equation}
		
		\item If the bistatic signal is undetected, the receiver relies only on passive bearing information from both sensor nodes.
		\begin{equation}\label{FIM_2}
			\mathcal{I}_{\bm{\theta}}  =  \tilde{\mathcal{I}}^{(S_1,k-1)}_{\bm{\theta}} +\tilde{\mathcal{I}}^{(S_2,k-1)}_{\bm{\theta}}
		\end{equation}
	\end{itemize}
	
	\subsection{Wideband Ambiguity Analysis}
	The CRLB offers a localized bound on estimation accuracy near the true parameter values. 
	To complement the CRLB analysis, we evaluate the WBAF, which characterizes a waveform’s ability to resolve time delay and Doppler scaling, two critical parameters that influence the distinguishability of targets and the overall localization performance. The WBAF between a signal $ u(t)$ and its delayed and scaled version is defined as
	\begin{equation}\label{eq:WBAF}
		\chi(\tau, \eta) = \sqrt{\eta}\int_{-\infty}^{\infty} u(t) \, u^{*}\left( \eta(t - \tau) \right) \, dt
	\end{equation}
	where $\tau,\eta$ represent the time delay and Doppler scaling factor, respectively, and $u^{*}(t)$ denotes the complex conjugate of $u(t)$. 

	Having established the signal and noise models, along with the CRLB as the performance metric, we now describe the simulation setup used to evaluate localization accuracy under simplified underwater conditions.
	
	\section{Simulation Setup}
	We consider two sensor nodes $S_1$ and $S_2$, positioned at $\mathbf{p}_1=\begin{bmatrix} -500  & 0 \end{bmatrix}^\top$ and $\mathbf{p}_2=\begin{bmatrix} 0 & 500 \end{bmatrix}^\top$, respectively. Each sensor node is composed of $M=5$ sensors. The AUV is assumed to weigh $w=1$ ton and is moving with a speed of  $v=5$ m/s in a particular direction. The wind speed at the sea surface is $w_s=3$ m/s. The target reflective strength for the AUV is taken as $\mathrm{TS}=-16$ dB re m$^2$. The speed of sound is taken as $c=1500 \, \text{m/s}$. 
	
	\subsection{SNR Design}
	The parameters in \eqref{SNR_active} and \eqref{SNR_passive}  are selected to reflect typical acoustic propagation and noise conditions in the Baltic Sea. They are given as
	\begin{align}
		\mathrm{SL}_{\text{active}} & = 170.8 + 10\log(P) \\
		\mathrm{SL}_{\text{passive}}(f) &= 60\log(v) + 9\log(w) - 20\log(f) + 35  \\
		\mathrm{PL}_i(r) & = 17 \log(r_i) \\
		\mathrm{NL}(f) &= 35 + 24\log(1 + w_s) - 17\log(f) 
	\end{align}
	where the $\mathrm{SL}_{\text{passive}}(f)$, $\mathrm{NL}(f)$ represent the passive signal and the noise spectral power level, respectively. Further, $f$ is the listening frequency in kHz, $P$ is the transmitted power, and $r_i, \,\, \forall i=\{1,2\}$ are the distances between the $i^{\text{th}}$ node and the target in meters.
	
	\subsection{Evaluated Communication Signals}
	In the CRLB analysis, the SPFSK and MFSK communication signals presented in \cite{lidstrom2024high} and \cite{lidstrom2023evaluation}, respectively, are considered. The communication signals have been designed for underwater applications with limited spectral occupancy and to be robust under the degradation of inter-symbol interference (ISI) and inter-carrier interference (ICI). To achieve this, the signals are designed with different coding schemes, resulting in varying spectral efficiencies and throughput. The specific signal parameters used are summarized in Table~\ref{Table1}. The communication signals are transmitted with a carrier frequency of $f_c=6$ kHz. These signals have been shown to achieve a good communication link with $\mathrm{SNR}>0$ dB. Therefore, the transmitted power $P$ is set to match this lower SNR limit for the direct path. 
	\begin{table}
		\caption{Design parameters for SPFSK and MFSK signals}
		\label{Table1}
		\centering
		\renewcommand{\arraystretch}{1.2}
		\begin{tabular}{l|c|c}
			\toprule
			\textbf{Parameter} & \textbf{SPFSK} & \textbf{MFSK (R2-S)} \\
			\midrule
			Alphabet Dimension & 1 & 16 \\
			Frame length ($N_\text{frame}$) & 2048 & 128 \\
			$\epsilon$ & 0.2 & 0.01 \\
			Number of tones & 256 & 256 \\
			Signal Duration (s)&  0.38 & 1.27 \\
			$G_{\text{\O{}}}$ & 2 & 2 \\
			$N_b$ & 1024 & 64 \\
			Bandwidth (kHz) &  4 & 4 \\
			Spectral efficiency (bit/s/Hz) & 0.149 & 0.024 \\
			\bottomrule
		\end{tabular}
	\end{table}
	
	Based on this configuration, we compute CRLBs to quantify localization accuracy in position and Doppler. We also analyze the WBAF of these waveforms to evaluate their resolution in delay and Doppler. 
	
	\section{Results and Discussion}
\begin{figure}[ht]
		\centering
		\begin{subfigure}{0.48\textwidth}
			\includegraphics[width=1\textwidth,trim={0.1cm 0.7cm 0.5cm 1cm}, clip]{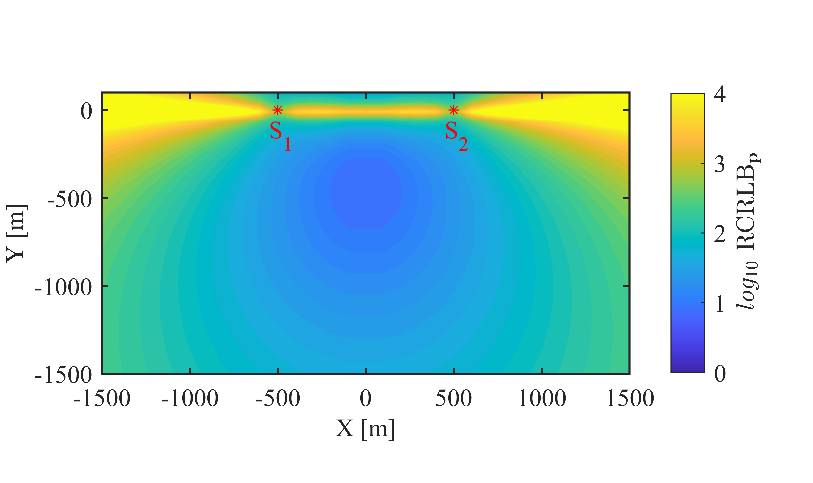}
			\subcaption{Passive Fusion: RCRLB for target position using only passive bearing measurements from sensor nodes $S_1$ and $S_2$.}
			\label{fig:fusion}
		\end{subfigure}
		\hfill
		\begin{subfigure}{0.48\textwidth}
			\includegraphics[width=1\textwidth,trim={0.1cm 0.3cm 0.5cm 0.1cm}, clip]{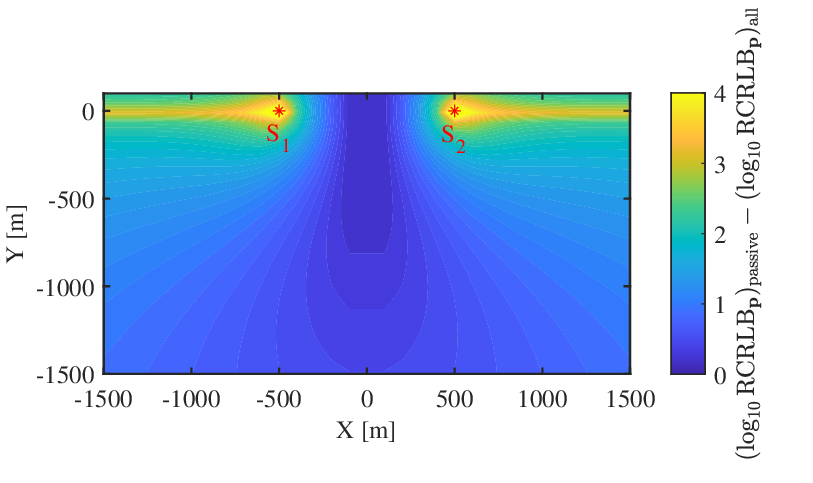}
			\caption{RCRLB improvement when bistatic measurements are combined with passive fusion using the SPFSK waveform.}
			\label{fig:vikmcook}
		\end{subfigure}
		\hfill
		\begin{subfigure}{0.48\textwidth}
			\includegraphics[width=1\textwidth,trim={0.1cm 0.3cm 0.5cm 0.1cm}, clip]{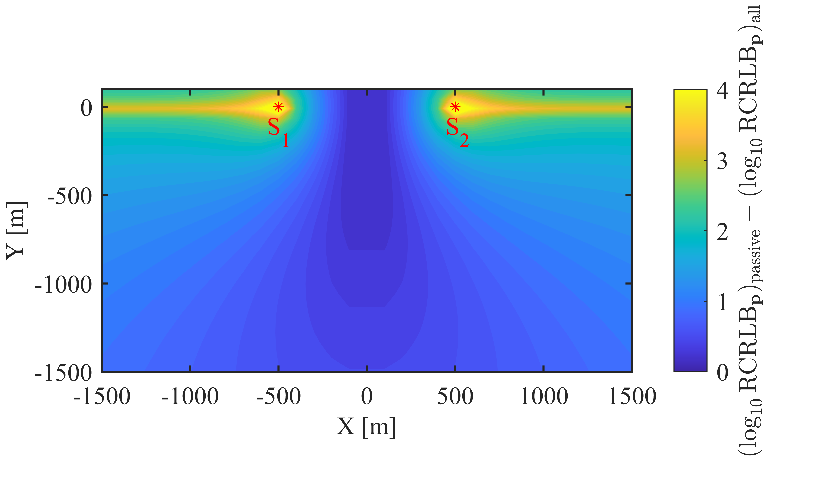}
			\caption{RCRLB improvement when bistatic measurements are combined with passive fusion using the MFSK waveform.}
			\label{fig:vik}
		\end{subfigure}
		\caption{Comparison of localization performance using RCRLB across different target positions under passive and bistatic measurement configurations.}
		\label{fig:crlb_all_pos}
	\end{figure}

	Fig.~\ref{fig:crlb_all_pos} illustrates how combining passive and bistatic measurements enhances localization accuracy, as reflected in RCRLB values. In particular, Fig.~\ref{fig:crlb_all_pos}(\subref{fig:fusion}) shows RCRLB values of approximately $10~\text{m}$ in the broadside region of both sensor nodes. For an AUV with an average length of 4 m, such localization accuracy may be acceptable. However, localization performance degrades significantly as the target moves toward the end-fire direction.
	\begin{figure}[ht]
		\centering
		\begin{subfigure}{0.48\textwidth}
			\includegraphics[width=1\textwidth]{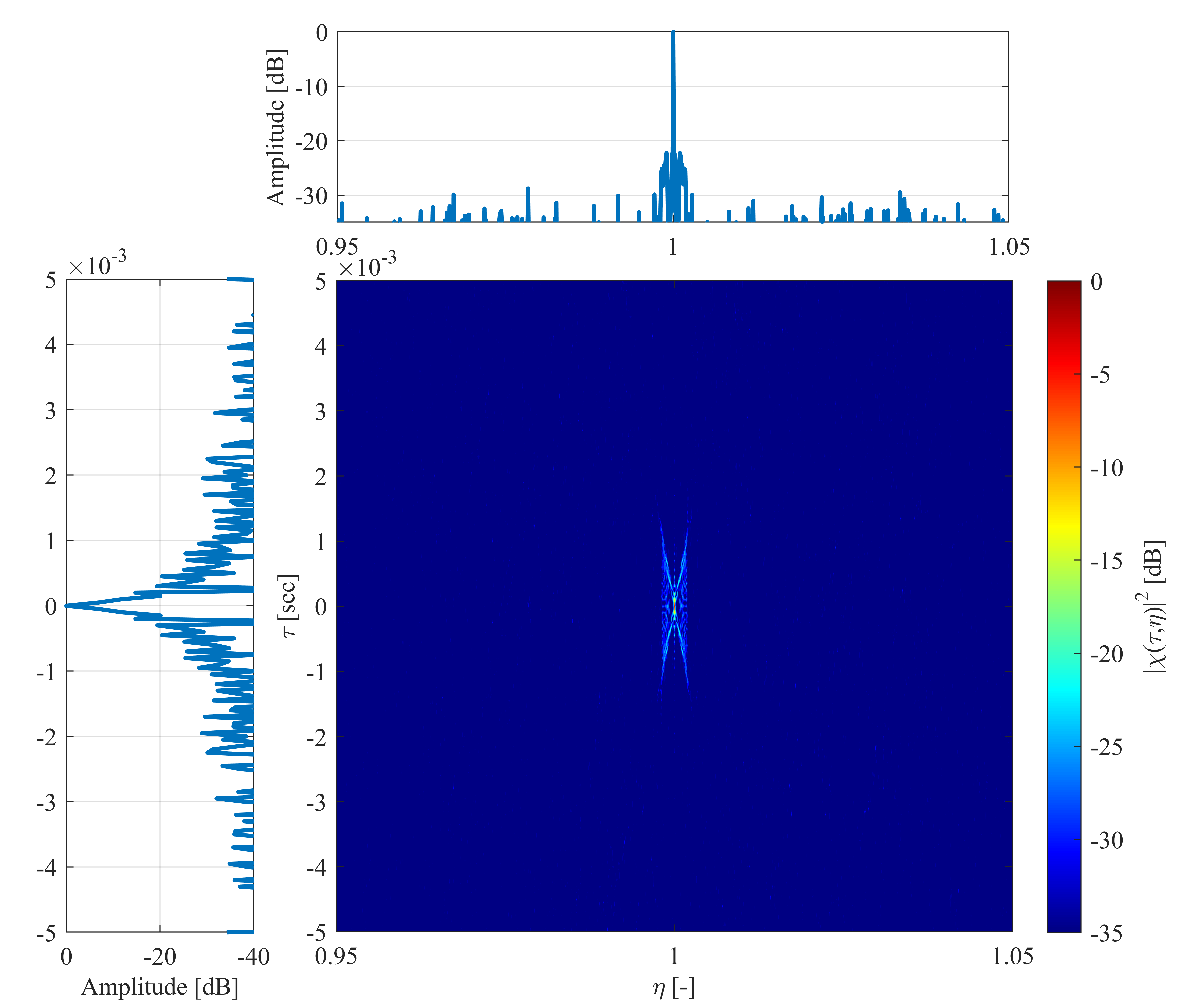}
			\caption{WBAF of SPFSK signal showing higher Doppler resolution and reduced sidelobe levels. Marginal plots depict delay and Doppler cross-sections.}
			\label{fig:WBAF_Mcook}
		\end{subfigure}
		\begin{subfigure}{0.48\textwidth}
			\includegraphics[width=1\textwidth]{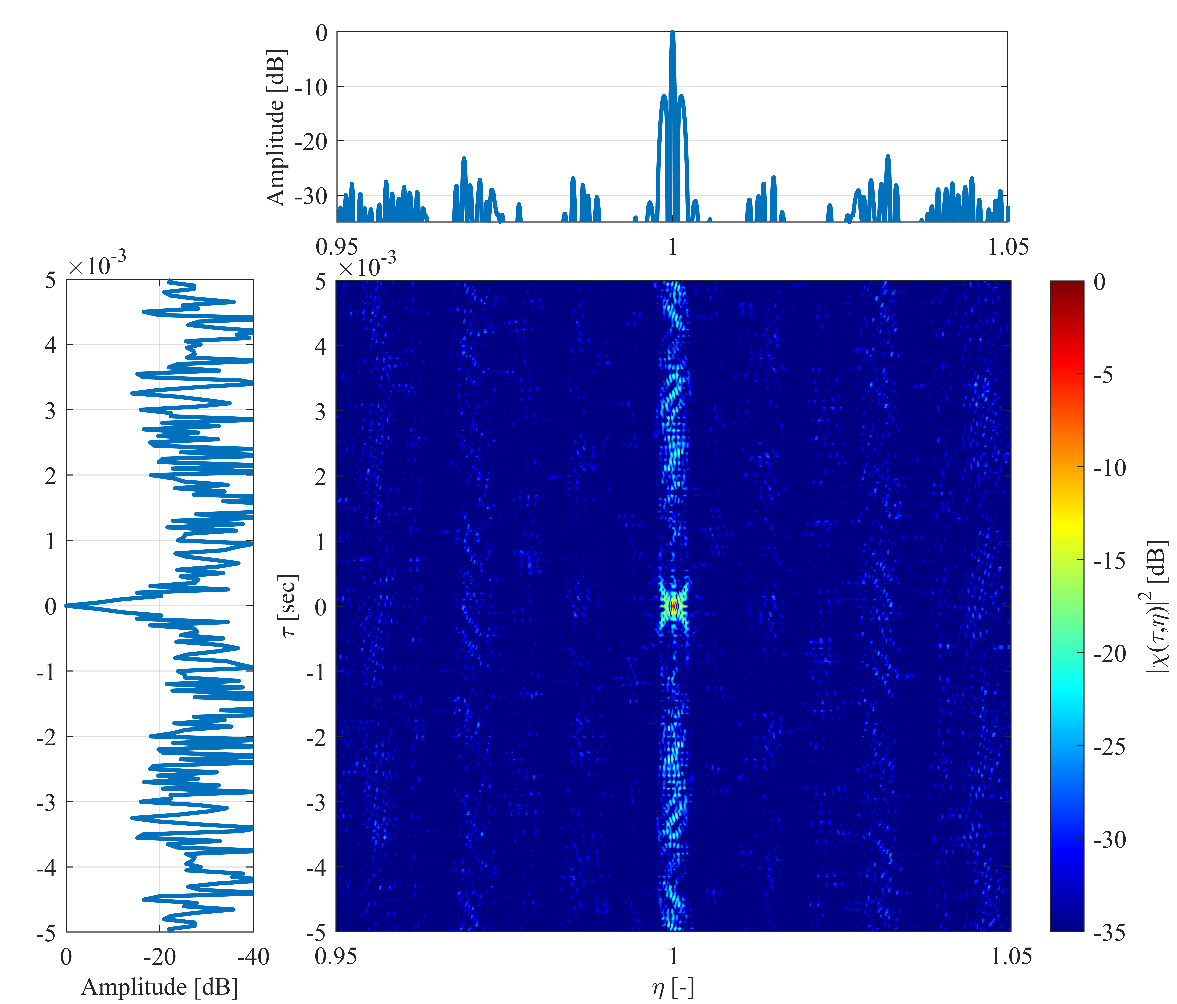}
			\caption{WBAF of MFSK signal showing comparable delay resolution but higher and more uniform sidelobe levels. Marginal plots depict delay and Doppler cross-sections.}
			\label{fig:WBAF_MFSK}
		\end{subfigure}
		\caption{Normalized WBAF plots for SPFSK and MFSK signals, illustrating their resolution capabilities in delay and Doppler domains.}
		\label{fig:WBAF}
	\end{figure}
	
	In contrast, Fig.~\ref{fig:crlb_all_pos}(\subref{fig:vikmcook}) and~\ref{fig:crlb_all_pos}(\subref{fig:vik}) demonstrate that incorporating bistatic information significantly enhances the accuracy across a wider spatial region. This improvement is indirectly influenced by the energy and occupied spectral bandwidth of the transmitted signal. Since both SPFSK and MFSK signals are designed to have identical transmitted energy and occupy the same bandwidth, the accuracy gains are comparable. Therefore, under equal bandwidth and energy constraints, the choice of waveform does not substantially impact localization accuracy.
	
	To evaluate the effectiveness of the communication signals as active bistatic sensing waveforms, we analyze their WBAF, shown in Fig.~\ref{fig:WBAF}. The SPFSK waveform exhibits superior Doppler resolution, attributed to its longer time duration, and shows significantly lower side-lobe levels up to -20 dB compared to MFSK. The delay resolution primarily depends on signal bandwidth and energy; thus, both signals demonstrate similar performance in the delay domain. However, their side-lobe behaviors differ notably. SPFSK displays a decaying side-lobe profile away from the main lobe, whereas MFSK maintains relatively constant side-lobe levels across delay offsets. These differences can be attributed to the inherent structural design of the two waveforms.
	\begin{figure}[t]
		\centering  
		\begin{subfigure}{0.48\textwidth}
			\includegraphics[width=1\textwidth,trim={0.1cm 0.1cm 0.5cm 0.1cm}, clip]{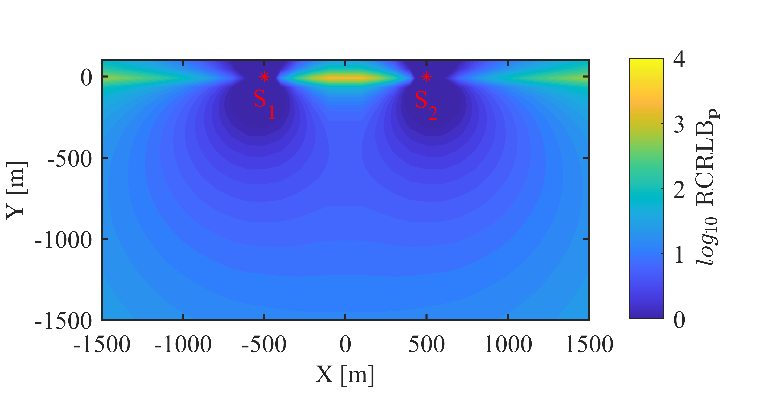}
			\caption{Position RCRLB using SPFSK waveform with both passive and active measurements.}
			\label{fig:spfsk_full_pos}
		\end{subfigure}
		\begin{subfigure}{0.48\textwidth}
			\includegraphics[width=1\textwidth,trim={0.1cm 0.1cm 0.5cm 0.1cm}, clip]{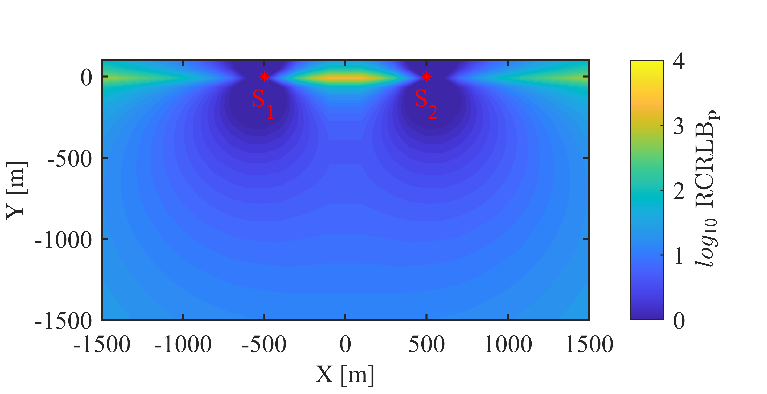}
			\caption{Position RCRLB using MFSK waveform with both passive and active measurements.}
			\label{fig:mfsk_full_pos}
		\end{subfigure}
		\hfill
		\begin{subfigure}{0.48\textwidth}
			\includegraphics[width=1\textwidth,trim={0.1cm 0.1cm 0.5cm 0.1cm}, clip]{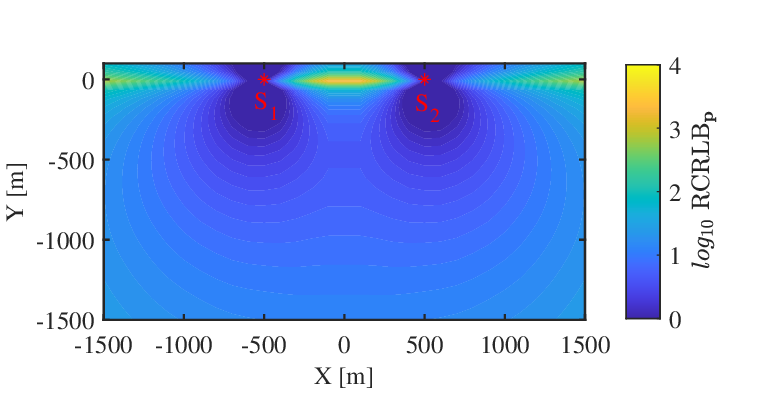}
			\caption{Position RCRLB using traditional LFM waveform, matched in bandwidth and duration to MFSK.}
			\label{fig:lfm_full_pos}
		\end{subfigure}
		\caption{RCRLB for target position using SPFSK, MFSK, and LFM waveforms under identical bandwidth constraints.}
		\label{fig:comparison_pos}
	\end{figure}
	
	Fig.~\ref{fig:comparison_pos} shows the comparison of the position RCRLB values of the communication signal with a traditional sonar signal. The LFM signal is designed to occupy the same bandwidth and time duration as that of the MFSK signal. We observe that with all the passive and active measurements, the RCRLB values range from approximately $4~\text{m}$ to $15~\text{m}$. These levels are achieved for a wider area around the sensor nodes. This shows that communication signals can achieve the same localization efficiency as that of traditional sensing signals.

\begin{figure}[t]
		\centering
		\begin{subfigure}{0.48\textwidth}
			\includegraphics[width=1\textwidth,trim={0.1cm 0.05cm 0.5cm 0.1cm}, clip]{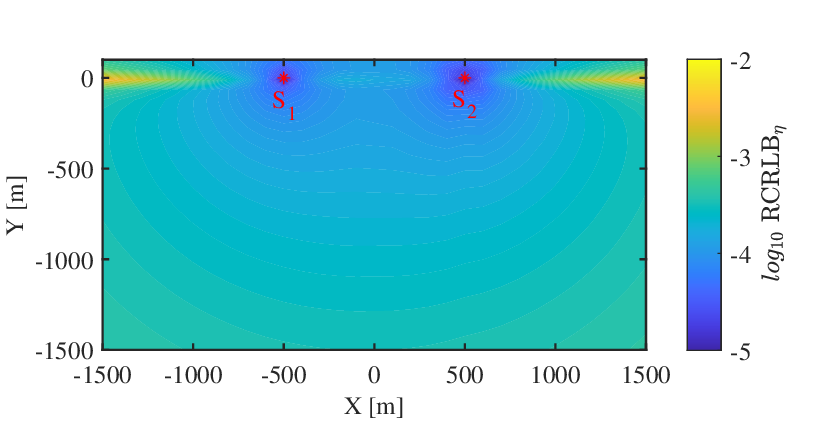}
			\caption{Doppler RCRLB using SPFSK waveform}
			\label{fig:dopler_vikmcook}
		\end{subfigure}
		\hfill
		\begin{subfigure}{0.48\textwidth}
			\includegraphics[width=1\textwidth,trim={0.1cm 0.05cm 0.5cm 0.1cm}, clip]{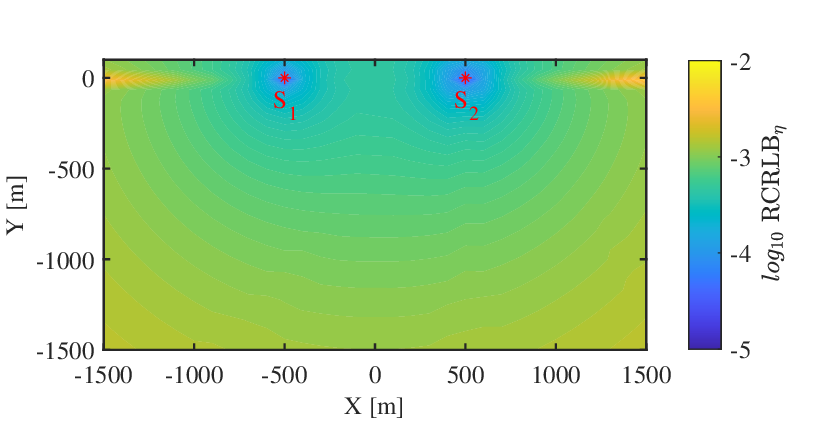}
			\caption{Doppler RCRLB using MFSK waveform.}
			\label{fig:dopler_MFSK}
		\end{subfigure}
		\begin{subfigure}{0.48\textwidth}
			\includegraphics[width=1\textwidth,trim={0.1cm 0.05cm 0.5cm 0.1cm}, clip]{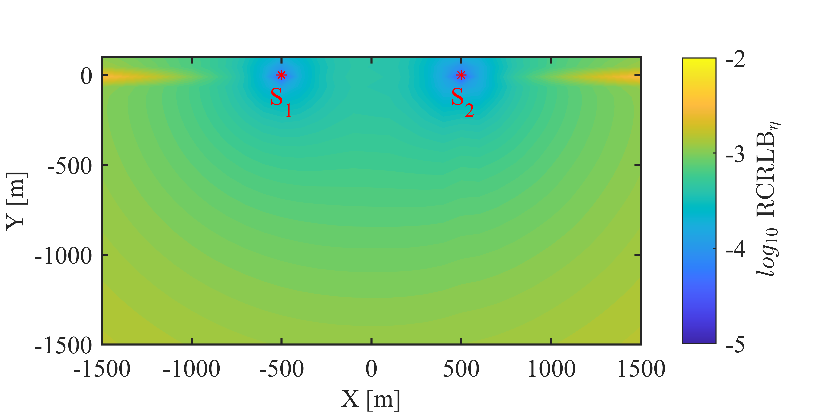}
			\caption{Doppler RCRLB using LFM waveform, showing similar performance to MFSK due to matched duration.}
			\label{fig:dopler_LFM}
		\end{subfigure}
		\caption{RCRLB for Doppler-scaling parameter $\eta$ using SPFSK, MFSK, and LFM waveforms, highlighting the impact of waveform duration on Doppler resolution.}
		\label{fig:doppler_all}
	\end{figure}

	Fig.~\ref{fig:doppler_all} shows the RCRLB for the Doppler-scaling parameter $\eta$, which reflects the accuracy in estimating the bistatic velocity of the target. Differences in RCRLB for Doppler estimation can be attributed to waveform duration, as longer signals offer finer Doppler resolution due to increased time-bandwidth product. The LFM Doppler performance is similar to that of the MFSK because of a similar time duration.

	These simulation results demonstrate that communication signals can effectively support target localization in underwater environments. Doppler effects arising from transmitter, receiver, and target motion, as well as multipath-induced delay spread, play a critical role in shaping signal design. These factors introduce trade-offs between spectral efficiency, throughput, and localization accuracy.

	\section{Conclusions and Future Work}\label{conclusion}
	In this paper, we examined the feasibility of using communication signals for localization within a bistatic underwater framework. Simulation results confirm that incorporating active bistatic information alongside passive bearing measurements significantly improves localization accuracy, as quantified by the position CRLB. Notably, this improvement is achieved without introducing additional sensing signals, highlighting the efficiency and practicality of ISAC systems for joint localization and communication tasks. Furthermore, we presented a CRLB-based framework that enables quantitative evaluation of waveform performance and supports the design of efficient ISAC systems. This model provides valuable insights into how waveform properties such as bandwidth and signal duration influence localization accuracy and inform the selection of signals suitable for underwater ISAC applications.
	
	This study is limited by simplified channel models and idealized assumptions. Future work should incorporate more realistic propagation effects, including multipath, reverberation, and synchronization errors, to validate the robustness of the proposed framework.
	
	\appendices
	\section{Details on the Signal Model}\label{appendix1}
	The signal model in~\eqref{sensor_mm}  can be elaborated by describing the signal received on the $m^{\text{th}}$ sensor of the $i^{\text{th}}$ sensor node during the $k^{\text{th}}$ data batch as
	\begin{align}\label{sensor_mm_full}
		y_{m,k}^{i}(n) & = u_k^{j,i}(\eta(\bm{\theta})(nT_s-\tau_\circ(\bm{\theta}) -\tau^{i}_m(\bm{\theta}))) \notag \\
		&\qquad  + s^{i}_{k}(nT_s-\tau^{i}_m(\bm{\theta})) +e_{m,k}^{i}(n)
	\end{align}
	Here, $n=1,\ldots,N_k$ and $T_s$ is the sampling time. \\
	The ambient noise $e_{m,k}^{{i}}(n)$ for the operational bandwidth follows the power law spectral behavior~\cite{Abraham2019} and is modeled using an autoregressive AR(1) process. Thus, the noise covariance for different time samples $(n,n')$ and sensors $(m,m')$ is given as  
	\begin{align}\label{noise_model}
		\mathbb{E}[e^{i}_{m,k}(n){e^{i}}_{m',k}(n')] &= \frac{\sigma_z^2}{1 - a^2} \left( a^{|n - n'|}\right) \delta_{m,m'}
	\end{align}
	where $a$ and $\sigma_z^2$ are the  AR(1) coefficient and variance of the innovation respectively.\\
	The signal parameters $\tau_\circ(\bm{\theta})$, $\tau^{i}_m(\bm{\theta})$ and $\eta(\bm{\theta})$ represent the bistatic time delay, inter-sensor delay for the $m^{\text{th}}$ sensor at the $i^{\text{th}}$ sensor node, and bistatic Doppler scaling, respectively. The signal parameters are dependent on $\bm{\theta} $ as follows
	\begin{align}
		\eta &= 1+\frac{1}{c}\left[\frac{\mathbf{v}\cdot (\mathbf{p}-\mathbf{p}_1)}{||\mathbf{p}-\mathbf{p}_1||} + \frac{\mathbf{v}\cdot (\mathbf{p}-\mathbf{p}_2)}{||\mathbf{p}-\mathbf{p}_2||}\right], \\   
		\tau_\circ & = \frac{1}{c} \Big[||\mathbf{p}-\mathbf{p}_1|| + ||\mathbf{p}-\mathbf{p}_2||\Big], \\
		\tau^{i}_m & =\frac{d}{c}(m-1)\sin(\psi_i),
	\end{align}   
	where $\mathbf{v}=\begin{bmatrix} v_x & v_y \end{bmatrix}^\top$ represents the target velocity, $\psi_i$ is the angle of the target with respective to negative y-axis and $\mathbf{p}_{i}=\begin{bmatrix}x_{i} & y_{i}\end{bmatrix}^\top$ represents the sensor nodes locations for ${i}=\{1,2\}$, $d$ and $c$ represent the inter-sensor spacing and the sound speed in the underwater environment respectively.\\
	Combining the signal received across all the sensors for the $k^{\text{th}}$ time slot at the $i^{\text{th}}$ sensor node, we can write
	\begin{equation}
		\mathbf{y}^{i}_{k}= \begin{bmatrix}
			\mathbf{y}^{i}_{1,k} \\
			\mathbf{y}^{i}_{2,k} \\
			\vdots \\
			\mathbf{y}^{i}_{M,k} \\
		\end{bmatrix}  \in \mathbb{R}^{MN_k \times 1}
	\end{equation}
	where
	\begin{equation}
		\mathbf{y}^{i}_{m,k}= \begin{bmatrix}
			y^{i}_{m,k}(1) \\
			y^{i}_{m,k}(2) \\
			\vdots \\
			y^{i}_{m,k}(N_k) \\
		\end{bmatrix}\in \mathbb{R}^{N_k \times 1} 
	\end{equation}
	The same structure is performed for the case when the signal $\mathbf{u}_k^{j, i}(\bm{\theta})$ is not available during the $k-1$ batch.

	\IEEEtriggeratref{7}
	\bibliographystyle{IEEEtran}
	\bibliography{IEEEabrv,OCEANS_paper_Great_lakes}

\begin{thebibliography}{10}
\providecommand{\url}[1]{#1}
\csname url@samestyle\endcsname
\providecommand{\newblock}{\relax}
\providecommand{\bibinfo}[2]{#2}
\providecommand{\BIBentrySTDinterwordspacing}{\spaceskip=0pt\relax}
\providecommand{\BIBentryALTinterwordstretchfactor}{4}
\providecommand{\BIBentryALTinterwordspacing}{\spaceskip=\fontdimen2\font plus
\BIBentryALTinterwordstretchfactor\fontdimen3\font minus
  \fontdimen4\font\relax}
\providecommand{\BIBforeignlanguage}[2]{{%
\expandafter\ifx\csname l@#1\endcsname\relax
\typeout{** WARNING: IEEEtran.bst: No hyphenation pattern has been}%
\typeout{** loaded for the language `#1'. Using the pattern for}%
\typeout{** the default language instead.}%
\else
\language=\csname l@#1\endcsname
\fi
#2}}
\providecommand{\BIBdecl}{\relax}
\BIBdecl

\bibitem{Bosser2025}
D.~Bossér, G.~Hendeby, M.~L. Nordenvaad, and I.~Skog, ``Broadband passive
  sonar track-before-detect using raw acoustic data,'' \emph{IEEE J. Oceanic
  Eng.}, pp. 1--11, 2025.

\bibitem{Akyildiz2005}
I.~F. Akyildiz, D.~Pompili, and T.~Melodia, ``Underwater acoustic sensor
  networks: research challenges,'' \emph{Ad Hoc Netw.}, vol.~3, no.~3, pp.
  257--279, May 2005.

\bibitem{Abraham2019}
D.~Abraham, \emph{Underwater {A}coustic {S}ignal {P}rocessing: {M}odeling,
  {D}etection, and {E}stimation}.\hskip 1em plus 0.5em minus 0.4em\relax
  Springer, 2019.

\bibitem{weiintegrated23}
Z.~Wei, H.~Qu, Y.~Wang, X.~Yuan, H.~Wu, Y.~Du, K.~Han, N.~Zhang, and Z.~Feng,
  ``Integrated sensing and communication signals toward {5G-A} and {6G}: A
  survey,'' \emph{IEEE Internet Things J.}, vol.~10, no.~13, pp.
  11\,068--11\,092, 2023.

\bibitem{liujrc20}
F.~Liu, C.~Masouros, A.~P. Petropulu, H.~Griffiths, and L.~Hanzo, ``Joint radar
  and communication design: Applications, state-of-the-art, and the road
  ahead,'' \emph{{IEEE} Trans. Commun.}, vol.~68, no.~6, pp. 3834--3862, 2020.

\bibitem{Keskin2024}
M.~F. Keskin, C.~Marcus, O.~Eriksson, A.~Alvarado, J.~Widmer, and H.~Wymeersch,
  ``Integrated sensing and communications with {MIMO}-{OTFS}: {ISI}/{ICI}
  exploitation and delay-doppler multiplexing,'' \emph{IEEE Trans. Wireless
  Commun.}, vol.~23, no.~8, pp. 10\,229--10\,246, Aug. 2024.

\bibitem{jun2018detection}
L.~Jun, Z.~Qunfei, Z.~Lingling, and S.~Wentao, ``Detection performance of
  active sonar based on underwater acoustic communication signals,'' in
  \emph{IEEE Int. Conf. on Signal Process., Commun. and Comput. (ICSPCC)},
  2018.

\bibitem{Men2022}
W.~Men, L.~Zhang, L.~Guo, and H.~Yin, ``Waveform design and processing for
  joint detection and communication based on {MIMO} sonar systems,'' in
  \emph{IEEE 5th Int. Conf. on Inf. Commun. and Sig. Proc. (ICICSP)}, Nov.
  2022, pp. 523--527.

\bibitem{niu2022integrated}
Q.~Niu, Q.~Zhang, and W.~Shi, ``Integrated waveform design scheme based on
  underwater detection and communication,'' in \emph{IEEE Int. Conf. on Signal
  Process., Commun. and Comput. (ICSPCC)}, Oct. 2022, pp. 1--5.

\bibitem{yin2020integrated}
J.~Yin, W.~Men, X.~Han, and L.~Guo, ``Integrated waveform for continuous active
  sonar detection and communication,'' \emph{IET Radar Sonar Navig.}, vol.~14,
  no.~9, pp. 1382--1390, 2020.

\bibitem{Wang2023}
J.~Wang, Y.~Cui, M.~Zhou, H.~Sun, L.~Liu, and A.~Zhang, ``Adaptive index
  modulated ofdm spread spectrum for underwater acoustic integrated sensing and
  communication networks,'' in \emph{IEEE Int. Conf. on Sig. Proc., Commun. and
  Comput. (ICSPCC)}, Nov. 2023, pp. 1--6.

\bibitem{Wang2024}
J.~Wang, J.~Lian, and G.~Zhu, ``A weighted mahalanobis distance target sensing
  strategy for underwater integrated sensing and communication systems,''
  \emph{IEEE Wireless Commun. Lett.}, vol.~13, no.~1, pp. 79--83, Jan. 2024.

\bibitem{Jehangir2024}
A.~Jehangir, S.~M. Majid~Ashraf, R.~Amin~Khalil, and N.~Saeed, ``{ISAC}-enabled
  underwater {IoT} network localization: Overcoming asynchrony, mobility, and
  stratification issues,'' \emph{IEEE Open J. of the Commun. Soc.}, vol.~5, pp.
  3277--3288, 2024.

\bibitem{Liu2016}
J.~Liu, Z.~Wang, J.-H. Cui, S.~Zhou, and B.~Yang, ``A joint time
  synchronization and localization design for mobile underwater sensor
  networks,'' \emph{IEEE Trans. Mob. Comput.}, vol.~15, no.~3, pp. 530--543,
  Mar. 2016.

\bibitem{KayI}
S.~M. Kay, \emph{Fundamentals of Statistical Signal Processing: {E}stimation
  Theory}.\hskip 1em plus 0.5em minus 0.4em\relax Prentice Hall, 1993.

\bibitem{Urick2013}
R.~J. Urick, \emph{Principles of underwater sound}, 3rd~ed.\hskip 1em plus
  0.5em minus 0.4em\relax Peninsula Publishing, Jun. 2013.

\bibitem{lidstrom2024high}
V.~Lidstr{\"o}m, ``High-speed noncoherent underwater acoustic communication
  using permutation alphabets and sliding hypothesis-tree decoding,'' \emph{J.
  Acoust. Soc. Am.}, vol. 155, no.~3, pp. 1840--1855, 2024.

\bibitem{lidstrom2023evaluation}
V.~Lidström, ``Evaluation of polar-coded noncoherent acoustic underwater
  communication,'' \emph{{IEEE} J. Ocean. Eng.}, vol.~50, no.~2, pp. 448--461,
  2025.

\end{thebibliography}
\end{document}